\documentclass[pra,aps,showpacs,preprint,unsortedaddress,superscriptaddress]{revtex4}
\usepackage{graphics,bm}
\usepackage{amssymb}
\usepackage{psfrag}
\usepackage{epsfig}
\usepackage{epsf}

\begin{document}

\title{Pinning and collective modes of a vortex lattice in a
Bose-Einstein condensate}

\author{J.W. Reijnders}
\email{jwr@science.uva.nl}
\affiliation{Institute for Theoretical Physics,
University of Amsterdam, Valkenierstraat 65, 1018 XE
Amsterdam, The Netherlands}

\author{R.A. Duine}
\email{duine@physics.utexas.edu}
\homepage{http://www.ph.utexas.edu/~duine} \affiliation{The
University of Texas at Austin, Department of Physics, 1 University
Station C1600, Austin, TX 78712-0264}

\date{\today}

\begin{abstract}
We consider the ground state of vortices in a rotating
Bose-Einstein condensate that is loaded in a co-rotating
two-dimensional optical lattice. Due to the competition between
vortex interactions and their potential energy, the vortices
arrange themselves in various patterns, depending on the strength
of the optical potential and the vortex density. We outline a
method to determine the phase diagram for arbitrary vortex filling
factor. Using this method, we discuss several filling factors
explicitly. For increasing strength of the optical lattice, the
system exhibits a transition from the unpinned hexagonal lattice
to a lattice structure where all the vortices are pinned by the
optical lattice. The geometry of this fully-pinned vortex lattice
depends on the filling factor and is either square or triangular.
For some filling factors there is an intermediate half-pinned
phase where only half of the vortices is pinned. We also consider
the case of a two-component Bose-Einstein condensate, where the
possible coexistence of the above-mentioned phases further
enriches the phase diagram. In addition, we calculate the
dispersion of the low-lying collective modes of the vortex lattice
and find that, depending on the structure of the ground state,
they can be gapped or gapless. Moreover, in the half-pinned and
fully-pinned phase, the collective mode dispersion is anisotropic.
Possible experiments to probe the collective mode spectrum, and in
particular the gap, are suggested.
\end{abstract}


\maketitle

\def\bx{{\bf x}}
\def\bk{{\bf k}}
\def\br{{\bf r}}
\def\bu{{\bf u}}
\def\half{\frac{1}{2}}
\def\args{(\bx,t)}

\section{Introduction} \label{sec:introduction}\
It has been known since the work of Onsager \cite{onsager1949} and
Feynman \cite{feynman1955} that a superfluid supports angular
momentum only through quantized vortices. Furthermore, following
Abrikosov's prediction that vortices in type-II superconductors
arrange themselves on a lattice \cite{abrikosov1957}, and its
experimental confirmation \cite{cribier1964,essmann1967},
Tkachenko showed that vortex lines in a rotating superfluid form a
regular hexagonal lattice in the absence of disorder
\cite{tkachenko1965}. Such an Abrikosov lattice, as it is nowadays
called, was indeed observed experimentally
\cite{tsakadze1978,yarmchuk1979}. Tkachenko also predicted the
vortex lattices to support phonons, the so-called Tkachenko modes
\cite{tkachenko1966}.

With the first experimental realization of Bose-Einstein
condensation in ultracold dilute atomic gases \cite{anderson1995},
another regime in the physics of neutral superfluids has become
accessible, i.e., the weakly-interacting regime. Following this
achievement, the same group created, for the first time, a vortex
in an atomic Bose-Einstein condensate \cite{matthews1999}.
Although there has been some experimental interest in the
equilibrium and nonequilibrium behavior of a single vortex line
\cite{anderson2000,rosenbusch2002,coddington2004}, since the
observation of a Bose-Einstein condensate with more than one
vortex \cite{madison2000}, however, most of the experimental
studies are focused on vortex lattices
\cite{raman2001,haljan2001,hodby2002}. In particular, the
dependence of the lowest Tkanchenko mode on the rotation frequency
has been measured \cite{coddington2003}, and is theoretically well
understood \cite{baym2003}.

One aspect that distinguishes the physics of vortices in atomic
Bose-Einstein condensates from superfluid Helium and
superconductors, is that in the latter systems the pinning of
vortices due to intrinsic disorder in the system plays an
important role
\cite{glaberson1966,weaver1972,fiory1978,baert1995,castellanos1997,
martin1997,morgan1998}. This, together with the discovery of
high-temperature superconductors, has led to many theoretical
studies of the effects of pinning on the melting of the vortex
lattice \cite{nelson1979,brezin1985,blatter1994,radzihovsky1995}.
Furthermore, in the context of type-II superconductivity, there
has been a lot of interest in the effects of a periodic array of
pinning centers on the ground state of the vortices
\cite{reichhardt1997,kolton1999,dasgupta2002,zhuravlev2003,pogosov2003}.
In particular, it turns out that, due to the competition between
vortex interactions and pinning, the system exhibits a rich ground
state phase diagram, as a function of the vortex density and the
strength of the pinning potential \cite{pogosov2003}. However,
since the pinning potential in the case of vortices in type-II
superconductors is known only phenomenologically, a detailed
comparison between theory and experiment seems unfeasible.

Very recently, we have shown that a rotating Bose-Einstein
condensate in a so-called optical lattice is a very attractive
system to study the pinning of vortex lattices in a superfluid
\cite{reijnders2004}. Such an optical lattice is formed by laser
fields that trap the atoms using the dipole force. Recently, the
experimental control over the strength of the optical lattice
enabled Greiner {\it et al.} \cite{greiner2002} to experimentally
explore the Mott-Insulator to superfluid quantum phase transition
\cite{fischer1989,jaksch1998}. By rotating the optical lattice at
the same frequency of rotation as the Bose-Einstein condensate,
the vortices experience a static pinning potential that is
determined by the optical lattice
\cite{reijnders2004,kevrekidis2003,bhattacherjee2003}. Such a
co-rotating optical lattice can be made by rotating holographic
phase plates or amplitude masks
\cite{boiron1998,dumke2002,chen2002,newell2003}. Since the
strength of the optical lattice determines the strength of the
pinning potential, and the rotation frequency controls the density
of vortices, the phase diagram can be studied in detail
experimentally.

In Ref.~\cite{reijnders2004} we have calculated the phase diagram
for a homogenous Bose-Einstein condensate with one vortex per unit
cell of the optical lattice analytically, by means of a
variational method. It is the aim of this paper to extend these
calculations to other vortex filling factors and to the situation
of a two-component Bose-Einstein condensate. Furthermore, we also
study the collective modes of the pinned and unpinned vortex
lattices. Complementary to our analytical work, Pu {\it et al.}
\cite{pu2004} numerically studied a Bose-Einstein condensate in a
co-rotating optical lattice with an additional harmonic confining
potential. The harmonic trapping potential leads to finite-size
effects which further enrich the phase diagram of the system.
Unfortunately, including an additional harmonic potential in our
variational calculations makes analytical results unfeasible.
Therefore, we consider the homogeneous case, which brings out the
physics of the competition between vortex interactions and pinning
potential most clearly. In Ref.~\cite{reijnders2004} we studied
both the case of a one-dimensional optical lattice and the
two-dimensional case. In this paper we focus on the
two-dimensional situation.

The paper is organized as follows. In Sec.~\ref{sec:pots} we
derive the pinning potential and vortex interaction energy. Using
these results, we calculate in Sec.~\ref{sec:energyvortexlattice}
the energy of an arbitrary vortex lattice in a periodic potential.
This result is used to determine the ground state phase diagram in
Sec.~\ref{sec:phasediag} for a single-component Bose-Einstein
condensate for various filling factors. The two-component case is
discussed in Sec.~\ref{sec:twocomp}. In Sec.~\ref{sec:collmodes}
we determine the dispersion of the low-lying collective modes over
the ground state. We end in Sec.~\ref{sec:concl} with our
conclusions.

\section{Vortex interactions and potential energy}
\label{sec:pots}

In this section we calculate the interaction energy of two
vortices, as well as the potential energy of a vortex in the
optical lattice, i.e., the pinning potential, by means of a
variational {\it ansatz}. These results are needed later on to
determine the phase diagram.

\subsection{Pinning potential}
\label{subsec:pinningpot} Since we assume the system to be at zero
temperature throughout the paper, the most convenient starting
point is the hamiltonian functional which gives the total energy
of the system in terms of the macroscopic condensate wave function
$\Psi (\bx)$, and reads
\begin{equation}
H[\Psi^*,\Psi]= \int {d \bf x}\Psi^*({\bf x}){\bigg [}
-{\hbar^2\nabla^2 \over 2 M} + {1\over2}g|\Psi({\bf x})|^2 +
V_{\rm OL}({\bf x})-\mu {\bigg ]}\Psi({\bf x})~.
\label{eq:hamiltonian}
\end{equation}
Here, $M$ denotes the mass of one atom which interacts with the
other atoms via a two-body contact interaction of strength $g=4
\pi a_s \hbar^2/M$, with $a_s>0$ the $s$-wave scattering length.
The two-dimensional optical lattice potential is given by
\begin{equation}
\label{eq:optpot}
  V_{\rm OL}({\bf x})=s E_{\rm R}\left[\sin^2(qx)+\sin^2(q y)\right]~,
\end{equation}
with $E_{\rm R}$ the recoil energy, $q$ the wavenumber of the
optical lattice, and $s\geq 0$ a dimensionless number indicating
the strength of the optical lattice. The chemical potential that
fixes the number of atoms in the condensate is given by $\mu$.

Throughout this paper we consider for simplicity a condensate with
infinite extent in the $x$-$y$-plane which is tightly confined in
the $z$-direction by an harmonic trap with frequency $\omega_z$.
This approach is motivated by the fact that a Bose-Einstein
condensate that is rotated around the $z$-axis will extend in the
$x$-$y$-plane due to the centrifugal force. Assuming that modes in
the $z$-direction are frozen out, such that the wave function
 is gaussian in this direction, effectively leads to a condensate thickness
$d_z\equiv\sqrt{\pi \hbar/(M\omega_z)}$. These assumptions allow
us to neglect the curvature of the vortex lines along the
$z$-direction. Note also that we can safely omit the term
proportional to the external rotation frequency in
Eq.~(\ref{eq:hamiltonian}), since we intend to work with a
variational ansatz which has a fixed vortex density, and,
moreover, we assume that the harmonic magnetic trapping potential
approximately cancels the centrifugal force.

We consider the system in the Thomas-Fermi limit where the kinetic
energy of the condensate atoms is neglected with respect to their
potential energy and mean-field interaction energy. Minimizing the
hamiltonian of Eq.~(\ref{eq:hamiltonian}) in this limit, the
global density profile of the condensate without vortices is given
by
\begin{equation}
 n_{\rm TF}({\bf x})=\left| \Psi (\bx) \right|^2=n-[V_{\rm OL}({\bf x})
-sE_{\rm R}]/g~,
\end{equation}
with $n=[\mu-sE_{\rm R}]/g$ the average density of the condensate.

As already mentioned, to find the potential energy of a vortex in
a Bose-Einstein condensate in an optical lattice, as a function of
its coordinates ${(u_x,u_y)}$, we use a variational {\it ansatz}
for the condensate wave function. It is given by
\begin{equation}
\Psi ({\bf x}) = \sqrt{n_{\rm TF}({\bf x})}~\Theta {\big [}{|{\bf x}-
{\bf u}|/\xi}-1{\big ]}\exp[i \phi({\bf x},{\bf u})]~,
\label{eq:vortexansatz}
\end{equation}
with $\xi=1/\sqrt{8 \pi a_s n}$ the healing length that sets the
size of the vortex core, $\phi({\bf x},{\bf
u})=\arctan[(y-u_y)/(x-u_x)]$ the phase configuration
corresponding to one vortex, and $\Theta(z)$ the unit step
function. For the above {\it ansatz} to be a good approximation,
we have assumed that the vortex core is much smaller then an
optical lattice period, $q \xi \ll 1$, and that the strength of
the potential is sufficiently weak, $s E_{\rm R} < \mu$. The use
of a unit step function for the density profile of the vortices is
justified because the main contribution to the energy of the
vortices is due to the superfluid velocity pattern and not due to
the inhomogeneity of the condensate density \cite{fetter2001}.

Substituting the {\it ansatz} in Eq.~(\ref{eq:vortexansatz}) in
the hamiltonian in Eq.~(\ref{eq:hamiltonian}) and integrating over
the entire $x$-$y$-plane gives the total energy of the vortex in
the optical lattice. This energy diverges with the system size.
However, we need to isolate the finite, position dependent
contribution to the energy due to the presence of the vortex,
which is the only relevant contribution for our purposes.

There are two position dependent terms which contribute
significantly to the energy. The first one is largest and is
entirely due to the kinetic energy of the condensate. Neglecting
the effect of the laplacian on the global density profile, which
is consistent with the Thomas-Fermi limit, we have
\begin{equation}
 U_{\rm kin}(u_x,u_y) = -\frac{ d_z s E_{\rm R}}{8 a_s}
  \int\!dxdy\left[ \frac{\sin^2 (qx) +
\sin^2 (qy)}{(x-u_x)^2+(y-u_y)^2}\right]~.
\end{equation}
The integral can be done by shifting the integration variables to
$x=\rho \cos\theta + u_x$ and $y=\rho \sin\theta + u_y$. A little
algebra yields
\begin{eqnarray}
 U_{\rm kin}(u_x,u_y) &=& \frac{ d_z s E_{\rm R}}{8 a_s}
\left[\cos(2qu_x)+\cos(2qu_y)\right]~\int_0^{2\pi}d\theta\int_\xi^\infty
\frac{d\rho}{2\rho}\cos(2 q\rho\cos\theta)\nonumber\\
~~~~~&\quad& -\frac{ d_z s E_{\rm R}}{8
a_s}\left[\sin(2qu_x)+\sin(2qu_y)\right]
~\int_0^{2\pi}d\theta\int_\xi^\infty\frac{d\rho}{2\rho}\sin(2
q\rho\sin\theta)~. \label{eq:pinresult}
\end{eqnarray}
When integrated over polar angle, the second part on the righthand
side of this expression gives zero. The integral in the remaining
part can be further simplyfied by using a Jacobi expansion and
integrating out the polar angle
\begin{equation}
\sum_{n=-\infty}^\infty (-1)^n\int_\xi^\infty
\frac{d\rho}{2\rho}J_{2n}(2q\rho)\int_0^{2\pi} d\theta e^{2 i
n\theta}=\pi\int_\xi^\infty\frac{d\rho}{\rho} J_0(2\rho q)\equiv
\pi Q_{\rm kin}(q\xi)~, \label{eq:kinfactor}
\end{equation}
where $J_l$ is the $l$-th orther Bessel function of the first kind.

The second vortex position dependent contribution to the energy
comes solely from the core. Let us consider the energy
contribution
\begin{equation}
U=d_z \int d^2 x \Psi^\ast({\bf x})\left[ V_{\rm OL}({\bf x})+
{g\over 2}|\Psi({\bf x})|^2 -\mu\right]\Psi({\bf x})~.
\end{equation}
Alternatively, this term is written as $U=E_\infty-U_{\rm
core}({\bf u})$, where $E_\infty$ is a divergent constant equal to
the energy of the condensate without a vortex and $U_{\rm
core}({\bf u})$ the contribution of the region excluded by the
core of the vortex. Since the latter depends on the vortex
coordinates this contribution must be taken into account which
gives
\begin{eqnarray}
U_{\rm core}({\bf u})&=& -d_z \int_{\rm core} d^2 x
\left[{g\over2}n_{\rm TF}({\bf x})n_{\rm TF}({\bf x})
+(V_{\rm OL}({\bf x - u})-\mu)n_{\rm TF}({\bf x})\right]\nonumber\\
&=& {d_z\over 2g}\int_{\rm core}d^2 x [\mu-V_{\rm OL}({\bf x -
u})]^2
\nonumber\\
&\simeq&-{d_z \mu\over g}\int_{\rm core}d^2 x V_{\rm OL}({\bf x -
u}) +O(s^2E^2_{\rm R})~.
\end{eqnarray}
Performing the integral on a disk with radius $\xi$ we arrive at
the same form as in Eq.~(\ref{eq:pinresult}). The only difference
is the prefactor, which depends on $q\xi$,
\begin{equation}
Q_{\rm core}(q \xi)=\frac{J_1(2 q\xi)}{2 q\xi}~.
\label{eq:geom_core}
\end{equation}
Consistent with our previous remarks, this contribution of the
vortex core to the position dependent energy is smaller than the
kinetic energy contribution. It adds to the latter contribution
given in Eq.~(\ref{eq:kinfactor}) and hence we define $Q\equiv
Q_{\rm kin}+Q_{\rm core}$. Putting things together, the potential
energy of a vortex described by the {\it ansatz} of
Eq.~(\ref{eq:vortexansatz}) in a two dimensional optical lattice
is given by \cite{footnote1}
\begin{equation}
U_{\rm pin}({\bf u})={ d_z \over 8 a_s}s E_{\rm R} Q(q \xi)
[\cos(2q u_x)+\cos(2q u_y)]~.
\label{eq:potential}
\end{equation}
 It is clearly seen that the potential
energy is minimal if the vortices are located at the maxima of the
optical potential. This is expected, since at these maxima the
condensate density, and hence the kinetic energy associated with
the superfluid motion, is minimal. The expression in
Eq.~(\ref{eq:potential}) is regarded as a pinning potential
experienced by vortices in a condensate loaded in a optical
lattice.

\subsection{Vortex interactions}\label{subsec:interaction}
The interaction energy of two vortices must be known explicitly to
calculate the ground state structure of vortex lattices. We
calculate this interaction energy by using the following {\it
ansatz} for the condensate wave function
\begin{equation}
\Psi ({\bf x}) = \sqrt{n}~\Theta(R-|\bx|)~\Theta {\big [}{|{\bf
x}- {\bf u}|/\xi}-1{\big ]}~\Theta {\big [}{|{\bf x}+ {\bf
u}|/\xi}-1{\big ]}\exp[i \phi({\bf x},{\bf u})+i \phi({\bf
x},-{\bf u})]~. \label{eq:twovortexansatz}
\end{equation}
This form is a generalization of the {\it ansatz} in
Eq.~(\ref{eq:vortexansatz}) to the case of two vortices in a
disk-shaped condensate with radius $R$ and average density $n$,
oppositely displaced over a distance $|{\bf u}|$ from the origin.
The reason that we do not explicitly take into account the spatial
inhomogeneity of the condensate density due to the optical lattice
potential in the calculation of the vortex interaction energy, is
that most of the vortices in the vortex lattice are separated by
more than one optical lattice constant, such that the effect of a
spatially varying density profile on the vortex interactions is
averaged out. In the relevant limit where the healing length is
small compared to the system size, the only significant
contribution comes from the kinetic energy of the condensate. For
simplicity we place the vortices along the $x$-axis,
$(u_x,u_y)=({r \over 2},0)$, which leads for the energy of the
system to
\begin{eqnarray}
 V(r) &=& \frac{\hbar^2 d_z n}{2 M}
 \int_0^{2\pi}\!d\theta\int_0^{R}\!d\rho\rho \frac{-64 \rho^2}
 { r^4+16 \rho^4-8 r^2 \rho^2 \cos2\theta}\nonumber\\
&=&  128 \pi\frac{\hbar^2d_z n}{2 M}\int_0^{R}\!d\rho\rho \frac{
\rho^2}
 { 16 \rho^4-r^4}{\rm sgn}(4\rho^2-r^2)~.
\label{eq:vortinter1}
\end{eqnarray}
Here $\rho$ is the radial coordinate and $\theta$ is the polar
angle. The effect of the condensate density profile is
incorporated by simply excluding the contribution of the vortex
cores from the expression in Eq.~(\ref{eq:vortinter1}) such that
\begin{eqnarray}
 V(r) &=& \frac{64 \pi\hbar^2 d_z n}{ M}\left[
 -\int_0^{ (r-\xi)/2}\!d\rho\rho \frac{ \rho^2}
 {16 \rho^4-r^4}  +
 \int_{ (r+\xi)/2}^{R}\!d\rho\rho \frac{\rho^2}
 {16 \rho^4-r^4} \right]~\nonumber\\
&=& -\frac{\pi\hbar^2 d_z n}{M}\lim_{{\tilde R}\rightarrow
\infty}\log \left [\frac{16{\tilde r}^6-4{\tilde r}^4 + 4{\tilde
r}^2 -1}{{\tilde r}^4 (16{\tilde R}^4-{\tilde r}^4)}\right]~,
\label{eq:interactiondiv}
\end{eqnarray}
where we defined ${\tilde R}\equiv R/\xi$ and ${\tilde r}\equiv
r/\xi$ and also took the limit $\tilde R \to \infty$. The latter
result is divergent with increasing system size. The finite,
interaction energy of the two vortex configuration is isolated by
subtracting the divergent constant $\frac{\pi\hbar^2 d_z
n}{M}\lim_{{\tilde R}\rightarrow \infty}\log[1/{\tilde R}^4]$ from
the expression in Eq.~(\ref{eq:interactiondiv}) and evaluating the
limit ${\tilde R}\rightarrow \infty$. The resulting expression
does not depend on the system size and behaves like
\begin{equation}
V(r)=-\frac{2\pi \hbar^2 d_z n}{M}\log\left(\frac{r}{\xi}\right)~,
\label{eq:interaction}
\end{equation}
for $r \gg \xi$. This is the well-known long range interaction
potential experienced by singly quantized vortices in two
dimensions \cite{kleinertbook}.

In the next section we will use the results for the vortex pinning
potential and the vortex interaction energy to calculate the
energy of a lattice of vortices.

\section{Energy of a vortex lattice in a periodic potential}
\label{sec:energyvortexlattice} In principle, to calculate the
equilibrium positions of the vortices, we have to minimize the
total energy as a function of the coordinates of the vortices.
Clearly, for a large number of vortices this is unfeasible. It is
known, however, that in the limit of strong pinning, the vortices
form regular lattices
\cite{reichhardt1997,kolton1999,dasgupta2002,zhuravlev2003,pogosov2003}.
Therefore, to find the phase diagram of the system, we minimize
the energy of the system assuming that the vortices form a regular
lattice. This procedure neglects the fact that for small pinning
potential the hexagonal Abrikosov vortex lattice is slightly
distorted by the pinning potential \cite{pogosov2003}.

To carry out the above minimization procedure, it is easiest to
parameterize a unit cell of the vortex lattice for a given filling
factor $\nu$. The filling factor is defined as the number of
vortices per pinning center, i.e., per minima of the pinning
potential. In terms of the density of vortices it is equal to $\nu
= n_v a^2$, where $n_v$ is the two-dimensional density of vortices
that is set by the rotation frequency $\Omega$ as
$n_v=M\Omega/(\pi \hbar)$ \cite{fetter2001}, and $a=\pi/q$ is the
optical lattice constant.

\begin{figure}
\psfrag{f}{$\varphi$} \psfrag{l2}{$L_2$} \psfrag{l1}{$L_1$}
\psfrag{bma}{$a \alpha l m$} \psfrag{ma}{$l a$} \psfrag{la}{$m a$}
\psfrag{ala}{ $a \beta m l$}
\psfrag{a}{$a$}
\includegraphics[width=9cm]{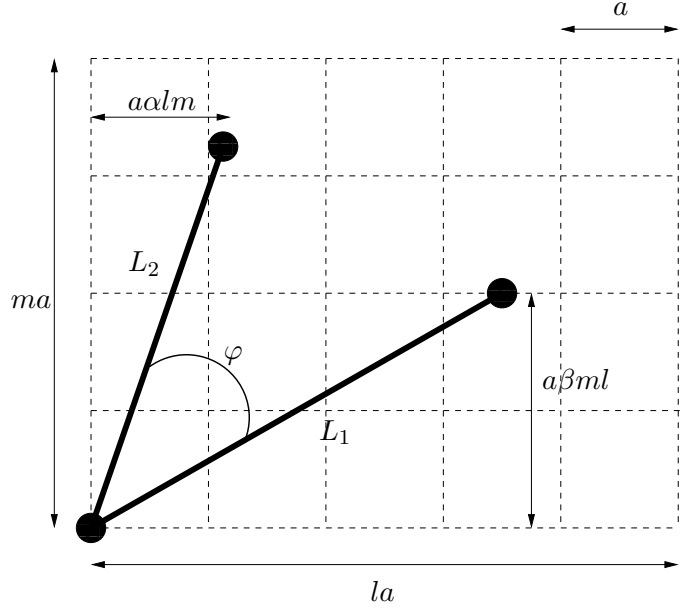}
\caption{\label{fig:unitcell} Two ways to parametrize the unit
cell of a vortex lattice, using parameters $(\alpha l, \beta m)$
and $(\varphi, \kappa)$, with $\kappa=L_1/L_2$. The relation
between those parametrizations is given by the expressions in
Eq.~(\ref{eq:relation}). The grid indicates the pinning potential
with lattice constant $a$.}
\end{figure}

We consider commensurate filling factors smaller than one, i.e,
$\nu = {1\over k}$, with $k$ a positive integer. All possible
vortex lattice unit cells corresponding to such commensurate
vortex lattices at a particular filling factor can be found by
factorizing $k$ in products $l\cdot m$, with $l$ and $m$ positive
integers, and arranging vortices on the sides of rectangles of
size $l a \times m a$, as shown in Fig.~\ref{fig:unitcell}.
Varying the vortex positions along the sides of the rectangle,
keeping the area of the unit cell constant, gives all possible
primitive commensurate lattice structures for the vortex lattice.
For a vortex lattice of filling $\nu$ this procedure is
parameterized by
\begin{equation}
{\bf u}(\alpha, \beta~;~l,m)~=~
a
\left(\begin{array}{cc}
\sqrt{1+\alpha l \beta m}&\alpha l\\ \beta m&\sqrt{1+\alpha l \beta m}
\end{array}\right)
\left(\begin{array}{c} l n_x\\m n_y
\end{array}\right)~,~
\label{eq:unitcellparametrization}
\end{equation}
with $n_i\in\mathbb{Z}$ and $0\leq \alpha,\beta\leq {1\over 2}$.
Notice that the transformation matrix in the above expression
preserves the area of the unit cell, since its determinant equals
unity. This ensures that we are considering lattice configurations
with equal vortex density. The more familiar parameters of a unit
cell of a two-dimensional lattice, the angle $\varphi$ between the
primitive lattice vectors and the ratio of their lengths,
$\kappa=L_1/L_2$, are related to $\alpha$ and $\beta$ by
\begin{eqnarray}
\frac{\cos \varphi}{\kappa}&=&\frac{m (\alpha l +\beta m)
\sqrt{1+\alpha l\beta m}}
{l[1+\beta m (\alpha l + \beta m)]}~,\nonumber \\
\frac{\sin \varphi}{\kappa}&=&\frac{m}{l[1+\beta m (\alpha l +
\beta m)]}~. \label{eq:relation}
\end{eqnarray}

The interaction energy $E_{\rm int}$ per unit cell as function of
$\varphi$ and $\kappa$ for an infinite two-dimensional lattice of
vortices subject to the logarithmic interaction potential of
Eq.~(\ref{eq:interaction}) was calculated by Campbell {\it et al.}
\cite{campbell1989}.  Cast in a dimensionless form their result
reads
\begin{eqnarray}
&& \tilde E_{\rm int} \equiv \frac{E_{\rm int}}{(\pi \hbar^2 d_z
n/M)} = \frac{\pi}{6} \frac{\sin \varphi}{\kappa}
 -\log \left[ 2 \pi \left( \frac{\sin \varphi}{\kappa
 }\right)^{1\over 2}\right]\nonumber \\
&&
 -\log \big\{\Pi_{j=1}^{\infty} [ 1-
 2 e^{-2\pi j|\sin\varphi|/\kappa}\cos\left(2 \pi j \frac{\cos \varphi}{
\kappa}\right)\nonumber\\
&& +~e^{-4\pi j |\sin \varphi|/\kappa}] \big\}~.
\label{eq:interactionenergy}
\end{eqnarray}
It is important to realize that the interaction energy per vortex
is divergent for an infinite vortex lattice, and that the above
expression gives the relative interaction energy for
configurations with equal vortex density. The absolute minimum of
the dimensionless interaction energy in
Eq.~(\ref{eq:interactionenergy}) corresponds to a hexagonal vortex
lattice structure, i.e., the Abrikosov vortex lattice with
$l=m=\sqrt{k}$ and $\alpha l=\beta m=\sqrt{1/\sqrt{3}-1/2}$ or
$(\varphi,\kappa)= (\pi/3,1)$, and is equal to ${\tilde E_{\rm
int}}=-1.32112$. Note that this lattice is incommensurate with the
optical lattice.

The pinning energy per unit cell is found by substituting
Eq.~(\ref{eq:unitcellparametrization}) in
Eq.~(\ref{eq:potential}), summing over all $n_x$ and $n_y$, and
dividing the result by the number of unit cells,
\begin{eqnarray}
E_{\rm
pin}(\alpha,\beta~;~l,m)&=&\lim_{P\rightarrow\infty}\frac{1}{4P^2}
\sum_{n_x=-P}^P\sum_{n_y=-P}^P U_{\rm pin}[{\bf u}(\alpha,
\beta~;~l,m)]
\nonumber\\
&=&-{ d_z \over 8 a_s}s E_{\rm R} Q(q \xi)
[\delta_{\beta m\in{\mathbb Z}}+\delta_{\alpha l\in{\mathbb Z}}]~.
\label{eq:pinningenergy}
\end{eqnarray}
This form of the pinning energy per unit cell is what we expect on
an infinite lattice. Only if the vortices form a lattice that is
commensurate with the optical lattice, they give a nonzero
contribution to the pinning energy. This is why we consider only
commensurate fillings, since we expect structural transitions at
these fillings. Incommensurate vortex lattices have zero potential
energy per unit cell on average. For $\nu \leq 1$ there are three
possible outcomes for the pinning energy in
Eq.~(\ref{eq:pinningenergy}). {\it i}) A phase in which all the
vortices are pinned by optical lattice maxima at $E_{\rm pin}=-{
d_z \over 4 a_s}s E_{\rm R} Q(q \xi)$, {\it ii}) a phase in which
a half of the vortices is pinned at $E_{\rm pin}= -{ d_z \over 8
a_s}s E_{\rm R} Q(q \xi)$, and {\it iii}) an unpinned phase at
$E_{\rm pin}=0$ for any vortex lattice that is incommensurate with
the optical lattice. The precise geometry of the unit cell of
these vortex lattices is determined further by minimization of the
interaction energy in Eq.~(\ref{eq:interactionenergy}). Of course,
the structure of the unpinned phase is always hexagonal,
corresponding to the global minimum of the interaction energy.

To end this section, we would like to point out that, since the
interaction energy of the vortex lattice is derived by summing the
expression for the interaction energy of two vortices over all
pairs of vortices, we have implicitly assumed that the vortex
density is so low that the vortex cores never overlap, and that we
are therefore allowed to neglect three-vortex interactions, and
interactions of higher order. A similar argument validates the
derivation of the pinning energy of the vortex lattice by summing
the single-vortex pinning potential over the number of vortices.
In the numerical calculations of Pu {\it et al.} \cite{pu2004},
these authors observed that for filling larger than one the
vortices form pinned phases where pairs of vortices are pinned,
and hence two vortices get very close together. Since our
approximations break down in this case, we study only phases with
a filling factor smaller than one.

\section{Phase diagrams} \label{sec:phasediag}
The energy per unit cell of the vortex lattice, obtained by adding
the pinning energy of Eq.~(\ref{eq:pinningenergy}) and the
interaction energy of Eq.~(\ref{eq:interactionenergy}), enables us
to calculate the zero-temperature phase diagram of the vortex
lattice structure at a certain filling. As already mentioned, we
consider systems with filling factor $\nu = {1 \over k}$ with $k$
a nonnegative integer larger than one.

The dimensionless energy per unit cell of the vortex lattice reads
\begin{equation}
\left(\frac{4 a_s}{\mu d_z}\right)E(\alpha,\beta~;~l,m)=
{\tilde E}_{\rm int}(\alpha,\beta~;~l,m)-{ 1 \over 2}{s E_{\rm R}\over \mu}
Q(q \xi)[\delta_{\beta m\in{\mathbb Z}}+\delta_{\alpha l\in{\mathbb Z}}]~,
\label{eq:dimensionless}
\end{equation}
where we used that $\mu\simeq g n $. It is most convenient to
minimize this expression in the plane spanned by the dimensionless
parameters $q \xi$ and ${s E_{\rm R}\over \mu}$. This leads to the
three phases discussed in the previous section. However, the
presence of the half-pinned vortex configuration depends on the
filling factor, implying different phase diagrams for even and odd
$k$. Since the structure of the lattice does not change
continuously, the phases are separated by a first-order
transition.

In the case of even $k$, the half-pinned lattice is absent, since
the pinning centers are distributed such that the minimum energy
configuration is always a fully-pinned lattice. The phase diagram
thus contains two distinct phases: a fully-pinned vortex lattice
and the hexagonal Abrikosov lattice. The geometry of the
fully-pinned vortex lattice is determined such that the
interaction energy is minimal
\cite{reichhardt1997,kolton1999,dasgupta2002,zhuravlev2003,pogosov2003}.

If $k$ is an odd integer, the half-pinned lattice is present in
the phase diagram if the pinning energy and the interaction energy
are of the same order. However, this phase exists only if the
inter-vortex distance and the optical lattice constant are
comparable in size.

\begin{figure}
\epsfig{file=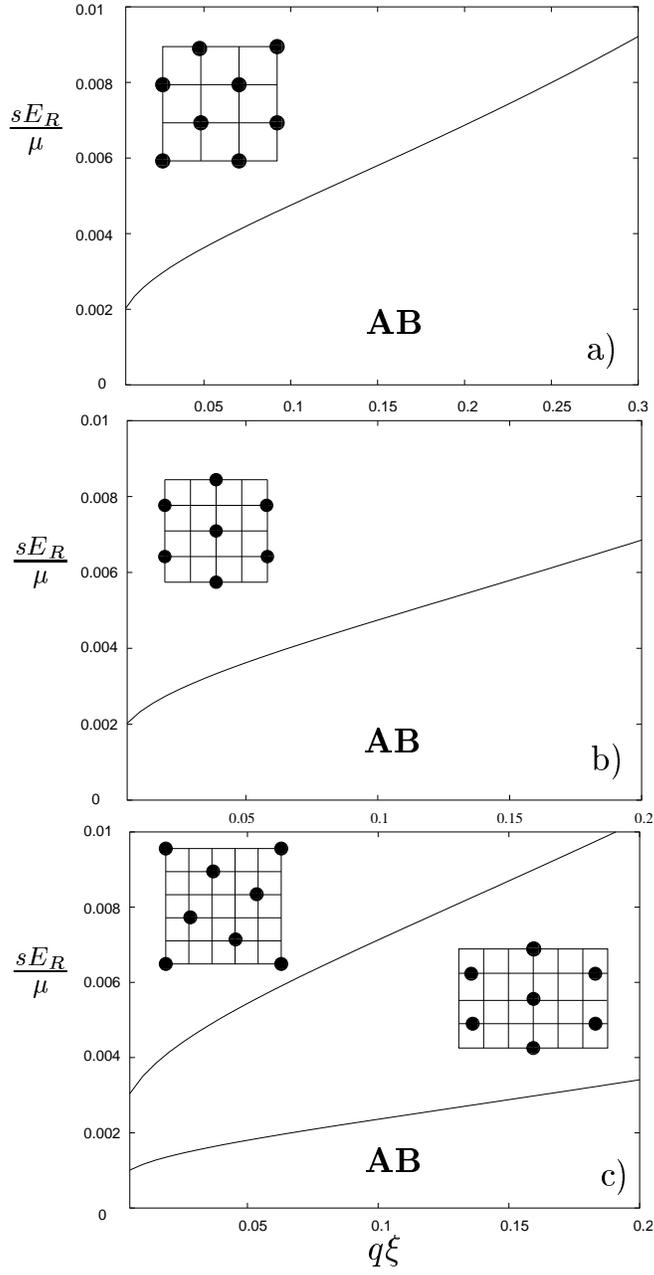,width=9cm} \caption{Vortex phase diagram of
a Bose-Einstein condensate in a two dimensional optical square
lattice, for three different filling factors, a) $\nu =
\frac{1}{2}$, b) $\nu=\frac{1}{4}$, and $\nu =\frac{1}{5}$. For a
weak pinning potential the vortex lattice structure is always the
hexagonal Abrikosov lattice (AB). The insets indicate the vortex
lattice structure for stronger pinning potential. The black dots
indicate the vortices, whereas the square grid indicates the
pinning potential. \label{fig:fig2}}
\end{figure}

In Ref.~\cite{reijnders2004} we discussed the case of one vortex
per optical lattice unit cell, i.e., $\nu =1$. We now discuss
three distinct examples in detail. The results are summarized in
Fig.~\ref{fig:fig2}.

\subsection{$\nu={1 \over 2}$} \label{subsec:half}
For $\nu=\frac{1}{2}$ the phase diagram contains two phases and is
depicted in Fig.~\ref{fig:fig2}~(a). For weak pinning the vortices
are not pinned and form a hexagonal Abrikosov lattice. For strong
pinning all vortices are located on the minimum of the pinning
potential, and form a square lattice with
$(\alpha,\beta)=(0,\frac{1}{2})$ and $(l,m)=(1,2)$. Note that, as
opposed to the $\nu=1$ case \cite{reijnders2004}, which also has a
square and pinned vortex lattice in the strong pinning regime, in
this case the vortex lattice is rotated over an angle
$\frac{\pi}{4}$ with respect to the optical lattice.

\subsection{$\nu={1 \over 4}$}\label{subsec:onefourth}
If there are four pinning centers per vortex, corresponding to
$k=4$, we find for large strength of the optical lattice a
fully-pinned triangular vortex lattice \cite{footnote2} with
$(\alpha,\beta)=(0,\frac{1}{4})$, $l=2$ and $m=2$. The interaction
energy per unit cell of the vortex lattice of this configuration
is ${\tilde E}_{\rm int}=-1.31849$. At small optical lattice
strength we find the hexagonal Abrikosov vortex lattice. The phase
boundary is given by
\begin{equation}
\left({s E_{\rm R}\over \mu}\right)_{\rm
hexagonal-pinned}={0.01057\over Q(q\xi)}~.
\label{eq:phaseboundaryfourth}
\end{equation}
It is important to note that, contrary to the case of
$\nu=\frac{1}{2}$ and $\nu=1$ \cite{reijnders2004}, the geometry
of the fully-pinned vortex lattice is in this case triangular.
Since a fully-pinned square lattice has the same pinning energy as
this triangular lattice, the interaction energy favors the latter.
The phase diagram for this filling is shown in
Fig.~\ref{fig:fig2}~(b).

\subsection{$\nu={1 \over 5}$}\label{subsec:onefifth}
At $k=5$ we find three phases. The result is depicted in
Fig.~\ref{fig:fig2}~(c). First, for large strength of the optical
lattice we have a fully-pinned vortex lattice with
$(\alpha,\beta)=({2\over 5},0)$, $l=5$, $m=1$ and interaction
energy ${\tilde E}_{\rm int}=-1.31055$. At intermediate optical
lattice strengths we find a half-pinned phase with
$(\alpha,\beta)=({1\over 2},0)$, $l=5$ and $m=1$. The interaction
energy per unit cell of this configuration equals ${\tilde E}_{\rm
int}=-1.31849$. At small lattice strength we find again the
hexagonal Abrikosov vortex lattice. The boundaries between these
phases are given by
\begin{equation}
\left({s E_{\rm R}\over \mu}\right)_{\rm hexagonal/
half-pinned}={0.00526\over Q(q\xi)}~,\quad \left({s E_{\rm R}\over
\mu}\right)_{\rm half-pinned/pinned}= {0.01588\over Q(q\xi)}~.
\label{eq:phaseboundaryfifth}
\end{equation}
Similar to the $\nu = \frac{1}{2}$ case. we find that the
fully-pinned vortex lattice has a square geometry, and is now
rotated over an angle $\tan^{-1}\left( \frac{1}{2}\right)$ with
respect to the optical lattice. Generally, if the fully-pinned
vortex lattice has a square geometry, then for filling factor
$\nu=\frac{1}{k_1^2+k_2^2}$, with $k_1$ and $k_2$ integer, the
fully-pinned vortex lattice will be rotated over an angle
$\tan^{-1} \left( \frac{k_2}{k_1}\right)$ with respect to the
optical lattice.

Contrary to the above mentioned filling factors, but similar to
the $\nu=1$ case \cite{reijnders2004}, there is an intermediate
triangular vortex lattice, where half of the vortices is pinned.

\section{Pinning of vortices in two-component condensates}\label{sec:twocomp}
In this section we study the influence of a two-dimensional
optical potential on vortex lattices in a mixture of Bose-Einstein
condensates of two different species. Our results also apply to a
Bose-Einstein condensate that consists of two hyperfine
components, provided the number of atoms in each component is
conserved. Along the lines of Sec. \ref{sec:pots},
\ref{sec:energyvortexlattice} and \ref{sec:phasediag} we calculate
the ground state phase diagram for two coupled condensates each
containing a vortex lattice at filling $\nu_i=1$ with the optical
potential. Note that the fact that we take the filling factor to
be the same in both species implies that the masses of both
species are approximately equal.

A system of two coupled Bose-Einstein condensates is described by
the following hamiltonian
\begin{eqnarray}
H&&=\sum_{i=1,2}\int {\textrm d}\mathbf{x}\Psi_i^\ast
\left[\frac{-\hbar^2 \nabla^2}{2 M_i}+V_{\rm{OL}} (\bx)
-\mu_i~\right]\Psi_i\nonumber\\
&&\quad\qquad\qquad\qquad +\int {\textrm
d}\mathbf{x}\left[\frac{1}{2}g_1|\Psi_1|^4+
\frac{1}{2}g_2|\Psi_2|^4+g_{12}|\Psi_1|^2|\Psi_2|^2\right],
\label{eq:two_cpt_ham}
\end{eqnarray}
with $g_i=4\pi\hbar^2 a_i/M_i$ and $g_{12}=2\pi\hbar^2
a_{12}/M_{ij}$. Here $a_{12}$ is the scattering length between
unlike species and the reduced mass is given by $M_{ij}=M_i M_j
/(M_i + M_j)$.

In the absence of the optical potential, Mueller and Ho
\cite{mueller2002} and Kasamatsu {\it et al}. \cite{kasamatsu2003}
theoretically predicted smooth transitions between hexagonal
lattices in both components at small (intra species) interactions
and interlaced square vortex lattices at larger interaction. These
square lattices where observed very recently by Schweikhard {\it
et al.} \cite{schweikhard2004}. However, the above-mentioned
transition is caused by the fact that the interaction energy is
minimized if the overall density is as smooth as possible. Since
we take a step function for the density profile of the vortex,
this density effect is not included in our calculations. Therefore
our results only make sense in the regime where this density
effect is dominated by the optical potential, i.e., in the
strong-pinning limit. To ensure this, the requirement
$\left(\frac{s E_R}{\mu_i} \right)Q(q\xi_i) \gg\frac{g_{12}}{g_i}$
must be satisfied. This implies that we must have that $g_i \gg
g_{12}$, since we assumed that $s E_R < \mu_i$. It must be
stressed that this is quite restricting as at the present day
there is no experimental atomic system known which meet these
requirements. However, one might expect that near an interspecies
Feshbach resonance this regime of parameters is realizable.
Therefore, we study the fully-pinned and half-pinned lattices and
the phase transition between them.

A non-rotating two-component condensate phase separates if
$g_{12}>\sqrt{g_1 g_2}$ \cite{ho1996,esry1997}. The condensates
mutually exclude each other, even in the absence of external
potentials \cite{timmermans1998} or {\it with} rotation
\cite{kasamatsu2003}. The vortex ground state in the latter case
will not be given by a regular lattice, in general. In our
calculations we restrict ourself to the regime where the system
does not phase separate. In this regime, the coupling parameters
satisfy $g_i>0$ and $g_1 g_2>g_{12}^2$, which falls safely within
the approximation discussed above. We define the dimensionless
parameter $\chi^2\equiv g_{12}^2/g_1 g_2$, for which these
criteria imply $0<\chi^2<1$.

Solving the coupled equations for the condensate wave functions,
derived from the hamiltonian of Eq.~(\ref{eq:two_cpt_ham}) in the
Thomas-Fermi approximation, leads to the following density profile
in component $i$
\begin{equation}
n_{\rm{TF}}^i(\mathbf{r})=|\Psi_i|^2=\frac{1}{g_i}
\frac{[\mu_i-V_{\rm{OL}}(\mathbf{r})]}{1-\chi^2} +\frac{1}{g_{12}}
\frac{[\mu_j-V_{\rm{OL}}(\mathbf{r})]} {1-\frac{1}{\chi^2}}~,\quad
j\neq i
\end{equation}
We use the variational {\it ansatz} for the condensate wave
function containing a vortex in component $i$
\begin{equation}
\Psi_i ({\bf x}) = \sqrt{n_{\rm TF}^i({\bf x})}~\Theta {\big
[}{|{\bf x}- {\bf u}|/\xi_i}-1{\big ]}\exp[i \phi_i({\bf x},{\bf
u})]~, \label{eq:twocptvortexansatz}
\end{equation}
with $\xi_i=1/\sqrt{8 \pi a_i n_i}$ and $n_i=(\mu_i-s E_{\rm
R})/g_i$. Furthermore, we assume that vortices in different
components do not interact. As explained before, in particular we
neglect the effect of the density profile caused by a vortex in
one component on the vortices in the other component, which, in
the absence of an optical potential, leads to the structural
transitions discussed by Mueller and Ho \cite{mueller2002} and
Kasamatsu {\it et al.} \cite{kasamatsu2003}. Within each
component, the vortex interactions are logarithmic, as derived in
Sec.~\ref{subsec:interaction}.

The pinning potential which is experienced by the vortex can be
calculated along the lines of Sec. \ref{subsec:pinningpot}. The
first contribution, coming from the kinetic energy term of the
hamiltonian in Eq.~(\ref{eq:two_cpt_ham}), is equal to
\begin{equation}
U^i_{\rm kin}({\mathbf u})=\frac{d_z}{8 a_i}s E_{\rm R} Q_{\rm
kin}(q \xi_i) G_i(g_1,g_2,g_{12})\left[\cos(2q u_x)+\cos(2 q
u_y)\right]~, \label{eq:twocptpotkin}
\end{equation}
with $Q_{\rm kin}$ given by Eq.~(\ref{eq:kinfactor}). The
difference with the single component case is the appearance of the
factor
\begin{equation}
G_i(g_1,g_2,g_{12})=\frac{g_{12}-g_i\chi^2}{g_{12}(1-\chi^2)}~.
\end{equation}
This factor is completely dependent on the various interaction
strengths. The other significant (position-dependent)
contribution, coming from the vortex core, involves more work,
\begin{eqnarray}
U^i_{\rm core}({\mathbf u})&=&-d_z\int_{\rm core}{\rm d}^2 x\left[
\frac{1}{2}g_i|\Psi_i|^4+(V_{\rm OL}-\mu_i)|\Psi_i|^2
+\frac{1}{2}g_{12}|\Psi_1|^2|\Psi_2|^2 \right]\nonumber\\
&=&-d_z\int_{\rm core}{\rm d}^2 x\left\{\frac{g_i}{2}\left[
\frac{(\mu_i-V_{\rm OL})^2}{g_i^2(1-\chi^2)^2}
+\frac{(\mu_j-V_{\rm OL})^2}{g_{12}^2(1-1/\chi^2)^2}\right.\right.\nonumber\\
&&\qquad\qquad\qquad\qquad\qquad\qquad\qquad\qquad\left.\left.
+2\frac{(\mu_i-V_{\rm OL})(\mu_j-V_{\rm OL})}{g_i g_{12}
(1-\chi^2)(1-1/\chi^2)}\right]\right.\nonumber\\
&+&\left[V_{\rm OL}-\mu_i +\frac{g_{12}}{2}\left(
\frac{(\mu_j-V_{\rm OL})}{g_j(1-\chi^2)} +\frac{(\mu_i-V_{\rm
OL})}{g_{12}(1-1/\chi^2)}
\right)\right]\nonumber\\
&&\qquad\qquad\qquad\qquad\qquad\qquad\qquad\times\left.\left[
\frac{(\mu_i-V_{\rm OL})}{g_i(1-\chi^2)}
+\frac{(\mu_j-V_{\rm OL})}{g_{12}(1-1/\chi^2)} \right]\right\}\nonumber\\
&=&-\frac{d_z}{g_i}\frac{1}{\chi^2-1}
\left[\frac{g_{12}(\mu_i+\mu_j)}{2 g_j}-\mu_i\right] \int_{\rm
core}{\rm d}^2 xV_{\rm OL}({\bf x}-{\bf u})+{\rm
O}(s^2E_R^2)~.\nonumber\\&&
\end{eqnarray}
If we assume that $\mu_i\approx \mu_j$ we find a contribution
similar to Eq.~(\ref{eq:twocptpotkin}). The prefactor in this case
is given by $Q_{\rm core}(q\xi)$ which is defined in
Eq.~(\ref{eq:geom_core}). The total pinning potential experienced
by a vortex in component $i$ due to the optical lattice is equal
to
\begin{equation}
U^i_{\rm pin} ({\mathbf u})=\frac{d_z}{8 a_i}s E_{\rm R} [Q_{\rm
kin}(q \xi_i) +Q_{\rm core}(q\xi_i)]
G_i(g_1,g_2,g_{12})\left[\cos(2q u_x)+\cos(2 q u_y)\right].
\end{equation}
This energy is dependent on the ratio of the coupling parameters.
The pinning energy is minimized on lattice maxima in both
components in the regime where it is the dominant energy scale.
Therefore, vortices in both components tend to be on the same
position.

\begin{figure}
\psfrag{g}{$\gamma\rightarrow$}\psfrag{se}{$\frac{s E_{\rm
R}}{\mu}$}
\begin{center}
\epsfig{file=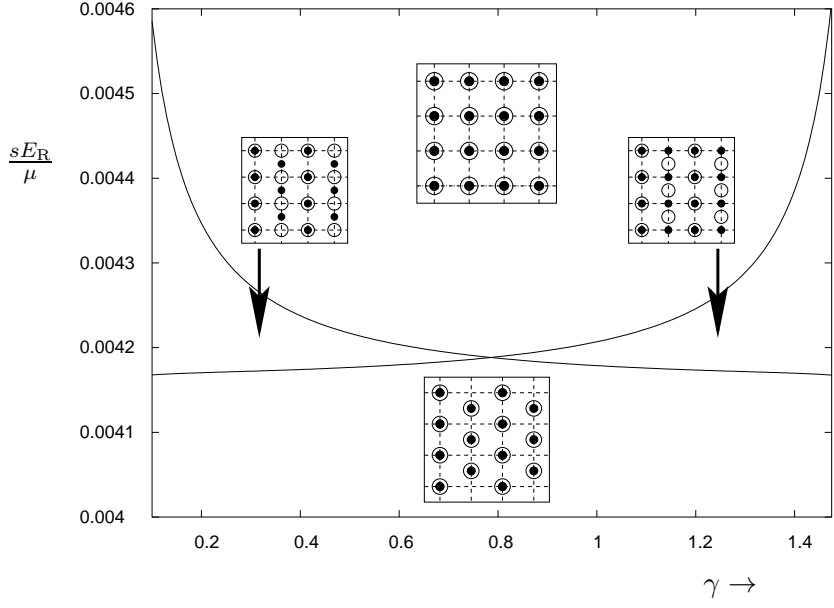,width=11cm}
\end{center}
\caption{\label{fig:twocptphasediagram} Vortex phase diagram of a
rotating two-component condensate in the presence of an optical
lattice, at commensurate filling, $\nu_i=1$. The phase diagram is
plotted as function of the dimensionless parameter
$\gamma=\arctan(g_1/g_2)$. Black dots and circles represent
vortices in different components. The dashed lines represent the
pinning potential. We took $\kappa=100$ and $q\xi=0.05$ which
implies $Q(0.05)\approx 3.612$.}
\end{figure}

In order to find the vortex phase diagram we minimize the
interaction and pinning energy in each component, as in Sec.
\ref{sec:phasediag}. We assume (again) $\mu_i\approx \mu_j=\mu$.
This implies $\xi_i\approx\xi_j=\xi$. As mentioned before, we are
only interested in lattice types which are fully-pinned or
half-pinned. The phase boundary between the fully-pinned and
half-pinned vortex lattices (commensurable with the optical
lattice) in each component is given by
\begin{equation}
\left(\frac{s E_{\rm R}}{\mu}\right)_i=
\frac{0.01494}{Q(q\xi)G_i(g_1,g_2,g_{12})}~.
\end{equation}
To find all possible lattice types in the two-component system it
is most convenient to parametrize the coupling parameters by
$g_1=\kappa g_{12}\sin(\gamma)$ and $g_2=\kappa
g_{12}\cos(\gamma)$. By varying $\gamma$ one scans along an circle
segment in the $(g_1,g_2)$-plane. The non-phase separated regime
in terms of the new parameters is given by $\kappa>\sqrt{2}$ and
$\gamma_-<\gamma<\gamma_+$ with
\begin{equation}
\gamma_\pm = \arccos\left[\sqrt{\half \left(
1\pm\frac{\sqrt{\kappa^4-4}}{\kappa^2}\right)}\right]~.
\end{equation}
We find that the  phase diagram contains four different vortex
lattices. In Fig. (\ref{fig:twocptphasediagram}) the phase diagram
and the lattice geometry are displayed for $q\xi=0.05$ and
$\kappa=100$.
Notice that the two-component phase diagram has a straightforward
interpretation in terms of coexistence of phases found in the
single-component case. For strong pinning the vortex lattices are
both fully pinned. For the filling under consideration ($\nu=1$)
the vortices in both components therefore form a square lattice.
Depending on the relative strength of the interaction, determined
by the parameter $\gamma$, the vortex lattice in one of the
components will change first to the half-pinned triangular
geometry, as one lowers the strength of the optical potential. For
sufficiently weak pinning potential both vortex lattices assume
this structure.

\section{Collective modes} \label{sec:collmodes}
In this section we calculate the dispersion of the collective
modes of the pinned vortex lattices. In principle, this requires
the calculation of the energy of the system for small
displacements ${\bf u} ({\bf r}_i)$ of the vortices from their
equilibrium positions ${\bf r}_i \equiv r_{ix} \hat x + r_{iy}
\hat y$. In first instance, the energy of the system is then given
by
\begin{eqnarray}
\label{eq:microscopicenergydistortedlattice}
  E \left[ {\bf u} \right] &=& \frac{d_z s E_{\rm R} Q(q\xi)}{8 a_s}
  \sum_i \left\{ \cos \left[ 2 q (r_{ix}+u_x ({\bf r}_i))\right] +
                 \cos \left[ 2 q (r_{iy}+u_y ({\bf
                 r}_i))\right]\right\} \nonumber \\
  &-& \frac{\pi \hbar^2  d_z  n}{M} \sum_{i \neq j} \log
   \left| \frac{{\bf r}_i+{\bf u} ({\bf r}_i) -{\bf r}_j - {\bf u} ({\bf r}_j)}{\xi}
   \right|~.
\end{eqnarray}
We perform a Fourier transform
\begin{equation}
  \bu (\br_i) = \frac{1}{\sqrt{N_x N_y}} \sum_{\bk} \bu (\bk)
   e^{i \bk \cdot \br_i}~,
\end{equation}
where $N_\alpha$ is the number of vortices along the $\alpha$
direction of the vortex lattice, and the momentum sum is
restricted over values $k_\alpha = 2\pi n_\alpha/L_\alpha$ in the
first Brillouin zone, where $n_\alpha$ is an integer and
$L_\alpha$ is the size of the vortex lattice in the $\alpha$
direction. Note that throughout this section we use Greek symbols
to indicate two-dimensional Cartesian components, i.e., $\alpha,
\beta  \in \left\{x,y \right\}$. We also sum over repeated Greek
indices.

We expect that after this Fourier transformation the energy of the
vortex lattice for small displacements will, up to an irrelevant
constant, be given by
\begin{equation}
\label{eq:energyvortexlatticeafterfourier}
  E \left[ \bu \right] = \frac{1}{2} \sum_\bk D_{\alpha \beta}
  (\bk)
  u_\alpha (-\bk) u_\beta (\bk)~,
\end{equation}
where the so-called dynamical matrix is the sum of three
contributions
\begin{equation}
\label{eq:dynmatrixsum} D_{\alpha \beta} (\bk) = D^{\rm
EL}_{\alpha \beta} (\bk) + D^{\rm LR}_{\alpha \beta} (\bk) +
D^{\rm OL}_{\alpha \beta} (\bk)~.
\end{equation}
The first contribution $D_{\alpha\beta}^{\rm EL} (\bk)$ takes into
account the interaction between neighboring vortices and follows
from elasticity theory. Although the elastic constants that will
enter in the expression for $D^{\rm EL}_{\alpha\beta} (\bk)$ can
in principle be calculated from the interaction energy in
Eq.~(\ref{eq:microscopicenergydistortedlattice}), such a
calculation is beyond the scope of this paper, and we will adopt a
more phenomenological point of view and write down the most
general form for $D^{\rm EL}_{\alpha\beta} (\bk)$ allowed by
symmetry arguments, for each lattice under consideration
\cite{landaulifschitzbook}. The second contribution
\begin{equation}
\label{eq:longrangedynmatrix}
  D^{\rm LR}_{\alpha\beta} (\bk) = \frac{8 \pi^2 \hbar^2  d_z  n}{A M}
\frac{k_\alpha k_\beta}{k^2+\lambda^{-2}}~,
\end{equation}
where $A$ is the area of the unit cell of the vortex lattice, is
independent of the structure of the lattice and follows from the
long-range nature of the logarithmic interactions, which has to be
taken into account separately
\cite{bonsall1977,abolfath2001,sinova2002}. Note that we have
explicitly included a finite range $\lambda$ of the logarithmic
interactions, to ensure that $D^{\rm LR}_{\alpha\beta} \to 0$ as
$k \to 0$. After taking the long wavelength limit, we can safely
take $\lambda \to 0$. The final contribution $D^{\rm
OL}_{\alpha\beta} (\bk)$ is due to the optical lattice.

The dispersion of the collective modes is determined by putting
the determinant of the matrix
\begin{equation}
\label{eq:matrixfordispersion}
  M_{\alpha\beta} (\bk,\omega)= D_{\alpha\beta} (\bk) -
  \epsilon_{\alpha \beta} 4\pi i n d_z \hbar \omega~,
\end{equation}
equal to zero. Here, $\epsilon_{\alpha \beta}$ is the
antisymmetric Levi-Cevita tensor in two-dimensions, that takes
into account the Euler dynamics of the vortices
\cite{fetter2001,kleinertbook}.

We will now calculate the dynamical matrix for each type of vortex
lattice considered in the previous section, i.e, for the
hexagonal, half-pinned, and fully-pinned vortex lattice, in the
long-wavelength limit, and use these results to calculate the
phonon spectrum of the vortex lattice.

\subsection{Hexagonal vortex lattice}
For a hexagonal vortex lattice we have that
\cite{landaulifschitzbook}
\begin{equation}
\label{eq:delabrikosov}
  D^{\rm EL}_{\alpha\beta}(\bk) = K_H k_\alpha k_\beta + \mu_H
  \delta_{\alpha \beta}  k^2~,
\end{equation}
where $K_H$ is the bulk modulus, and $\mu_H$ is the shear modulus
of the hexagonal vortex lattice. Using this result together with
Eq.~(\ref{eq:longrangedynmatrix}), we find for the frequency of
the collective modes in the absence of an optical lattice that
\begin{equation}
\label{eq:tkachenkofreq}
  \hbar \omega_\bk = \frac{k}{4\pi n d_z} \sqrt{\mu_H
  \left[ \frac{8 \pi^2 \hbar^2  d_z  n}{a_v^2 M}+\left(K_H+\mu_H
  \right)k^2\right]}~.
\end{equation}
For large wavelengths we have that $\hbar \omega_\bk \simeq c_T
k$, where the sound velocity of the so-called Tkachenko waves is
given by
\begin{equation}
\label{eq:soundvelocitytkachenko}
  c_T = \frac{1}{4\pi n d_z} \sqrt{
 \frac{8 \pi^2 \mu_H \hbar^2  d_z  n}{a_v^2 M}}~.
\end{equation}
We could now fix the value for the shear modulus of the vortex
lattice $\mu_H$ by demanding that the Tkachenko sound velocity is
equal to $\sqrt{\hbar \Omega/(4M)}$, the result known from the
hydrodynamic theory of a vortex lattice
\cite{tkachenko1966,sonin1987}. This would, however, not be
consistent, since the value for $\mu_H$ should follow from the
expression for the energy of the vortex lattice in
Eq.~(\ref{eq:microscopicenergydistortedlattice}), and may lead to
a different sound velocity due to the variational approximations
we have made in the description of the vortex lattice. Note that
is crucial to take into account the long-range interactions of the
vortices by means of the dynamical matrix in
Eq.~(\ref{eq:longrangedynmatrix}) to get a linear dispersion at
long wavelengths, since an omission of this part in the dynamical
matrix would lead to a quadratic dispersion. Moreover, we would
like to point out that due to the fact that vortices are described
by Euler dynamics we find only one mode, instead of two modes for
the case of a lattice of particles that obey Newtonian dynamics.

The polarization of the vortex lattice phonons is determined by
the eigenvector of the matrix in
Eq.~(\ref{eq:matrixfordispersion}), corresponding to the
eigenfrequency in Eq.~(\ref{eq:tkachenkofreq}). Generally, the
displacements are given by $\bu (\br_i,t) = \bu_{\bk,0} e^{i \bk
\cdot \br_i - i \omega_\bk t}$, where $\bu_{\bk,0}$ is the
eigenvector. For a wave in the $y$-direction we have that
\begin{eqnarray}
\label{eq:eigenvectortkachenko}
  \bu_{\bk,0} \propto
 \left( \begin{array}{c}
             \frac{i\left[\frac{8 \pi^2 \hbar^2  d_z  n}{a_v^2 M} +
k_y^2\right]}{k_y\sqrt{\mu_H
  \left[ \frac{8 \pi^2 \hbar^2  d_z  n}{a_v^2 M}+\left(K_H+\mu_H
  \right)k^2\right]}}\\
             1
         \end{array}
        \right)~
\approx \left( \begin{array}{c}
             \frac{i}{\mu_H k_y} \sqrt{\frac{8 \pi^2 \mu_H \hbar^2
d_z  n}{a_v^2 M}}\\
             1
         \end{array}
        \right)~,
\end{eqnarray}
which shows that the vortices move on an ellipse with the long
axis perpendicular to the direction of propagation. In the limit
$\bk \to 0$, the wave is almost transverse.

The translation symmetry of the system is broken explicitly in the
presence of an optical lattice. The collective modes therefore
acquire a gap, i.e., there is a minimum amount of energy required
to excite a phonon. Considering the part of the energy of
Eq.~(\ref{eq:microscopicenergydistortedlattice}) which corresponds
to the pinning energy of the vortices and expanding it, we have,
up to an irrelevant constant,
\begin{eqnarray}
E^{\rm OL} \left[ {\bf u} \right] = -2q^2\frac{d_z s E_{\rm R} Q(q\xi)}{8 a_s}
  &\sum_{\bk,\bk'}& \left\{u_x(\bk) u_x(\bk')\sum_j\cos[2 q r_{jx}]
e^{i( \bk + \bk') \cdot \br_j}\right.\nonumber\\
 ~& + & \left. u_y(\bk) u_y(\bk')\sum_j\cos[2q r_{jy}]
e^{i(\bk + \bk') \cdot \br_j}\right\},
\label{eq:hexOLenergy}
\end{eqnarray}
with the positions of vortices in the hexagonal lattice given by
\begin{equation}
\br_i=a_v \sqrt{\frac{2}{ \sqrt{3} } }\left[(i_x + \half i_y) \hat{x}
+\half \sqrt{3}i_y \hat{y}\right].
\end{equation}
Summing over these lattice coordinates in
Eq.~(\ref{eq:hexOLenergy}) gives zero. This is in contradiction
with the fact, that on symmetry grounds we expect a finite gap for
the collective modes. To overcome this problem we must take into
account that the real unpinned vortex ground state in a weak
periodic pinning potential is a slightly distorted hexagonal
lattice \cite{pogosov2003}. This comes about because the periodic
potential exerts a small net force on the vortices arranged in the
hexagonal lattice. The small distortion of the vortex positions is
a modulation around the equilibrium of the hexagonal lattice
$\br_i \mapsto \br_i + {\bf R}(\br_i)$. Following Pogosov {\it et
al}. \cite{pogosov2003} we find for a square two-dimensional
periodic potential
\begin{equation}
 R_\alpha(\br_i)=\frac{s E_{\rm R}}{2 q \mu_H A}\frac{d_z}{8 a_s}
Q(q \xi)\sin[2 q r_{i\alpha}]~.
 \label{eq:modulation}
\end{equation}
The modulation of the vortex coordinates around the positions of
the regular hexagonal lattice involves a factor $\eta=\frac{s E_{
\rm R}}{\mu_H A}$. To keep the displacements small, the condition
$\eta \ll 1$ must be satisfied.

Summing over the displaced hexagonal vortex lattice in
Eq.~(\ref{eq:hexOLenergy}) indeed leads to a non-zero energy gap.
Numerical evaluation of Eq.~(\ref{eq:hexOLenergy}) on a large
lattice gives a contribution to the dynamical matrix which is
roughly linear in $\eta$,
\begin{equation}
D_{\alpha\beta}^{\rm OL} \approx  0.5 \eta ~s E_{ \rm R}
\left[\frac{q d_z Q(q\xi)}{4 a_s}\right]^2\delta_{\alpha\beta}~,
\qquad\eta \ll 1~.
\end{equation}
 Including this in the calculation of the collective modes, we find
\begin{eqnarray}
\label{eq:tkachenkofreq_with_ol}
  \hbar \omega_\bk = && \frac{1}{4\pi n d_z} \left[0.5 \eta
~s E_{\rm R} \left(\frac{q d_z Q(q\xi)}{4 a_s}\right)^2+\mu_H
k^2\right]^\half
\times \nonumber\\
 && \left[ \frac{8 \pi^2 \hbar^2  d_z  n}{a_v^2 M}+0.5 \eta ~s E_{\rm
 R}
\left(\frac{q d_z Q(q\xi)}{4 a_s}\right)^2+\left(K_H+\mu_H
  \right)k^2\right]^\half ~.
\end{eqnarray}
 The gap takes the form
\begin{equation}
\hbar \omega_{\bf 0} \approx \frac{0.5 \eta ~s E_{\rm R}}{4\pi n
d_z} \left[\frac{q d_z Q(q\xi)}{4 a_s}\right]^2.
\end{equation}

\subsection{Half-pinned vortex lattice}
In the case of the half-pinned vortex lattice, which is always
triangular, the dynamical matrix that follows from elasticity
theory is given by \cite{landaulifschitzbook}
\begin{equation}
\label{eq:dynmatrixhalfpinned}
  D^{\rm EL}_{\alpha\beta} (\bk) = K_{T} k_\alpha k_\beta + \mu_T
  \delta_{\alpha\beta}k^2 + \kappa_T
  (1-\delta_{\alpha\beta})k_\alpha k_\beta + \delta\!\mu_T
  \tau^{z}_{\alpha \beta} k_\alpha k_\beta~,
\end{equation}
where $K_T$ is the bulk modulus of the triangular vortex lattice,
and $\kappa_T$ and $\mu_T$ denote the Lam\'e constants. For a
square lattice we have that the parameter $\delta\!\mu_T$ is equal
to zero, as we will see below. In the above expression
$\tau^z_{\alpha\beta}$ denotes the Pauli matrix.

To find the contribution due to the optical lattice, we consider
specifically the half-pinned case at filling $\nu=\frac{1}{5}$, as
shown in Fig.~\ref{fig:fig2}. We parametrize the equilibrium
position of the vortices as
\begin{equation}
\label{eq:parahp}
   \br_i = \frac{5}{2} i_x a \hat x + 2 \left[\frac{1-(-1)^{i_x}}{4} \right]
a \hat y + 2 i_y  a \hat y~.
\end{equation}
With the use of this parametrization we find that the energy of
the vortex lattice due to the pinning of the optical lattice is,
for small displacements $\bu (\br_i)$ from the equilibrium
positions, given by
\begin{eqnarray}
\label{eq:pinningenergyhalfpinned}
  E^{\rm OL} \left[ {\bf u} \right] &=& \frac{d_z s E_{\rm R} Q(q\xi)}{8 a_s}
  \sum_i \left\{ \cos \left[ 2 q (r_{ix}+u_x ({\bf r}_i))+\pi \right] +
                 \cos \left[ 2 q (r_{iy}+u_y ({\bf
                 r}_i))+\pi \right]\right\}  \nonumber \\
  &\simeq&
   \frac{d_z s E_{\rm R} q^2 Q(q\xi)}{4 a_s} \left[
    \sum_i (-1)^{i_x+1}  u_x^2 (\br_i) + u_y^2 (\br_i)
    \right]~,
\end{eqnarray}
where we have omitted an irrelevant constant. Note that we have
translated the optical lattice potential to ensure that there is a
vortex at the origin, consistent with the parametrization in
Eq.~(\ref{eq:parahp}). Using that
\begin{equation}
  \sum_i e^{i \bk \cdot \br_i} = N_x N_y \delta_{\bk,{\bf 0}}~,
\end{equation}
we find after a Fourier transform in first instance that
\begin{equation}
   E^{\rm OL} \left[ {\bf u} \right] =
   \frac{d_z s E_{\rm R} q^2 Q(q\xi)}{4 a_s}
    \sum_\bk \left[   \sum_i \sum_{\bk'} (-1)^{i_x} e^{i (\bk+\bk') \cdot \br_i}
      u_x (\bk) u_x (\bk') +
    u_y (\bk) u_y (-\bk) \right]~.
\end{equation}
The sum $i_x$ in the second term of this equation is evaluated by
splitting it into a sum over $i_x$ even and $i_x$ odd. If we
denote $\bk+\bk' = (2\pi n_x/L_x, 2 \pi n_y/L_y)$, we have that
\begin{eqnarray}
  && \sum_i (-1)^{i_x+1} e^{i (\bk+\bk') \cdot \br_i} =
  -\sum_{i_x,i_y} (-1)^{i_x} e^{2\pi i n_x i_x/N_x+ 2 \pi i n_y/N_y
  \left\{ \left[ \frac{1-(-1)^{i_x}}{4} +i_y \right]\right\}  }
  \nonumber \\
  &&=- N_y \delta_{n_y,0} \left[ \sum_p e^{2 \pi i n_x (2 p)/N_x} -
  \sum_l e^{2\pi i n_x (2p+1)/N_x - \pi i n_y/N_y} \right]
  \nonumber \\
  && =- N_x \delta_{n_x,0} N_y \delta_{n_y,0}
  \left[ \frac{1}{2} - \frac{1}{2} e^{2\pi i n_x/N_x-\pi i
  n_y/N_y}\right].
\end{eqnarray}
With the use of this result we have that
\begin{equation}
   E^{\rm OL} \left[ {\bf u} \right] =
   \frac{d_z s E_{\rm R} q^2 Q(q\xi)}{4 a_s}
    \sum_\bk
      u_y (\bk) u_y (-\bk) \equiv \frac{1}{2} \sum_\bk D_{\alpha\beta}^{\rm OL}
      (\bk) u_\alpha (\bk) u_\beta (-\bk)~,
\end{equation}
so that the contribution to the dynamical matrix due to the
optical lattice is given by \cite{footnote3}
\begin{equation}
 D^{\rm OL}_{\alpha\beta} (\bk) =
   \frac{d_z s E_{\rm R} q^2 Q(q\xi)}{2 a_s} \delta_{\alpha,y}
   \delta_{\alpha\beta}~.
\end{equation}

With the above results, we find that the collective mode
dispersion is given by
\begin{eqnarray}
\left(4\pi n d_z \hbar \omega_\bk \right)^2&=&
   \mu_T^2 k^4 +\mu_T^2 k^2 \left(K_T k^2 +
\frac{8 \pi^2 \hbar^2  d_z  n}{a_v^2 M} \right)
   -\kappa_T \left[\frac{16 \pi^2 \hbar^2  d_z  n}{a_v^2 M}+
(2K_T+\kappa_T)k^2\right] \frac{k_x^2 k_y^2}{k^2}
   \nonumber \\
   &-&
   \frac{1}{4} \delta\!\mu_T^2 k_x^2 k_y^2 + \mu_T \delta\!\mu_T k^2
   (k_x^2-k_y^2)\nonumber \\
   &+&\frac{d_z s E_{\rm R} q^2 Q(q\xi)}{2 a_s}
   \left[k_x^2 \left(K_T +\mu_T -\frac{\delta\!\mu_T}{2}+
\frac{8 \pi^2 \hbar^2  d_z  n}{a_v^2 M k^2} \right) +\mu_T
   k_y^2\right]~.
\end{eqnarray}
Interestingly, this dispersion is gapless, i.e., $\hbar
\omega_{\bf 0} =0$. The eigenvector corresponding to this
eigenfrequency is given by $(1,0)$, and so the displacement of the
vortices is along the $x$-axis. Physically, this is understood
because it does not cost any energy to uniformly displace all the
vortices in the $x$-direction when the vortices are forming a
half-pinned lattice with the geometry shown in
Fig.~\ref{fig:fig2}. (See again \cite{footnote2} and
\cite{footnote3}.) This comes about because under a uniform
translation of the vortices in the $x$-direction, half of the
vortices move away from an energy minimum and therefore increase
their energy, the other half moves downhill from an energy
saddle-point, precisely compensating this increase.

\subsection{Fully-pinned vortex lattice}
The structure of the fully pinned vortex lattice at certain
filling is triangular, in general. Therefore, the contribution to
the dynamical matrix due to the elasticity of the vortex lattice
takes the form of Eq.~(\ref{eq:dynmatrixhalfpinned}) with elastic
constants $K_P,~\mu_P~,\kappa_P$ and $\delta\!\mu_P$. In the
special cases for which the fully pinned vortex lattice has a
square structure, we have that $\delta\!\mu_P=0$. Since all the
vortices are positioned at the minimum of the pinning potential,
the contribution of the optical lattice to the dynamical matrix is
constant and diagonal, and given by
\begin{equation}
 D^{\rm OL}_{\alpha\beta} (\bk) =
   \frac{d_z s E_{\rm R} q^2 Q(q\xi)}{2 a_s}
   \delta_{\alpha\beta}~.
\end{equation}
Hence, we find for the phonon dispersion
\begin{eqnarray}
\left( 4 \pi n d_z \hbar \omega_\bk \right)^2 &=&  \left(
\mu_P k^2 + \frac{d_z s E_{\rm R} q^2 Q(q\xi)}{2 a_s}\right)^2
\nonumber\\
&&+ \left( \mu_P k^2 + \frac{d_z s E_{\rm R} q^2 Q(q\xi)}{2
a_s}\right) \left[K_P k^2+\frac{8 \pi^2 \hbar^2  d_z  n}{a_v^2 M
k^2}
+ \frac{\delta\!\mu_P}{2}\left(k_x^2-k_y^2\right)\right]\nonumber\\
&&+ \left[ \frac{\delta\!\mu_P^2}{4}-\kappa_P^2 - 2 \kappa_P
\left( K_P+\frac{8 \pi^2 \hbar^2  d_z  n}{a_v^2 M k^2}\right)
\right] k_x^2 k_y^2~.
\end{eqnarray}
At zero momentum we find for the gap \cite{footnote4}
\begin{equation}
  \hbar \omega_{\bf 0} = \frac{q^2 s E_{\rm R} Q (q\xi)}{8 n
  a_s}~,
\label{eq:pinnedgap}
\end{equation}
corresponding to an eigenvector $\propto (1,i)$. From this
eigenvector we therefore conclude that the zero-momentum mode
physically corresponds to a precession of the vortices around the
maxima of the optical lattice potential, as expected.

\subsection{The gap}
\begin{figure}
\psfrag{gap}{$\frac{\hbar\omega_0}{\mu}\times 10^{-5}$}
\psfrag{V}{$\frac{s E_{\rm R}}{\mu}$}
\epsfig{file=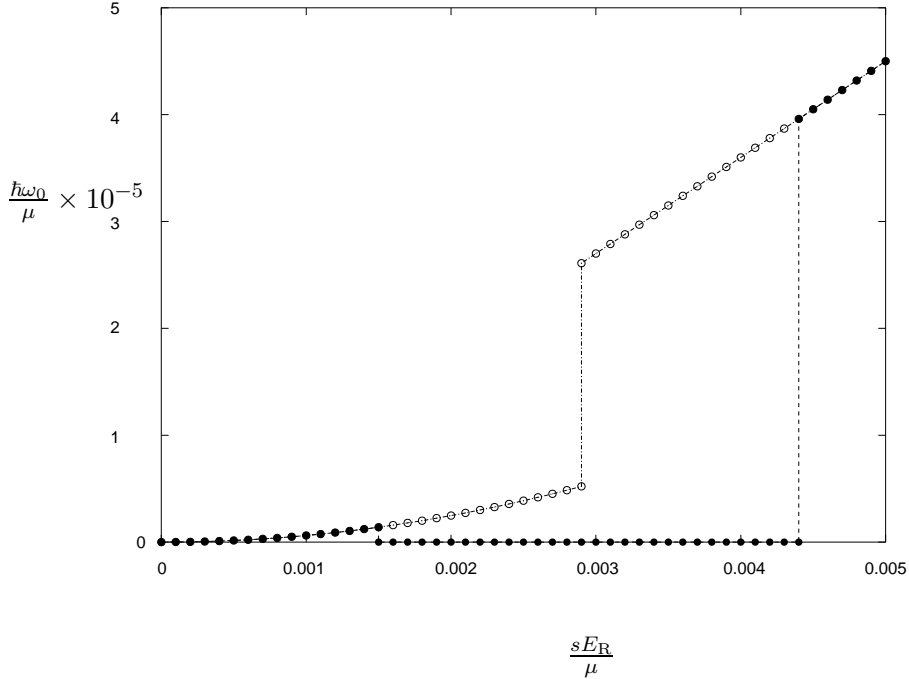,width=12cm} \caption{\label{fig:gap_vs_V}
The energy gap to the collective modes of a vortex lattice in the
presence of a periodic optical potential as function of the
lattice strength, for $\nu=\frac{1}{4}$ (open circles) and
$\nu=\frac{1}{5}$ (filled circles). We used $\eta=0.64
\left(\frac{s E_{\rm R}}{\mu}\right)$ (see \cite{footnote5}) and
$q\xi=.05$.}
\end{figure}

If one tunes the strength of the periodic potential, the vortex
lattice changes, depending on the filling $\nu$. The energy gap of
the collective modes is different  in the three vortex lattice
phases. To summarize the results on the gap, we have
\begin{equation}
\hbar \omega_{\bf 0} =\left\{
\begin{array}{ll}
0.5~ \eta~\frac{d_z}{8 a_s}[q\xi Q(q\xi)]^2s E_{\rm R} &
\textrm{hexagonal vortex lattice,}\\
 0 &
\textrm{half-pinned vortex lattice,}\\
 (q\xi)^2 Q(q\xi)s E_{\rm R}/\pi &
\textrm{fully-pinned vortex lattice.}
\end{array}
\right. \label{eq:resultsgap}
\end{equation}
It is clear that the gap in the hexagonal phase is much smaller
than the gap in the fully-pinned vortex phase. This is because in
the hexagonal phase, the non-zero contribution comes entirely from
the displacements of the vortices from the equilibrium positions
of the hexagonal lattice. The gap is then of second order in $s
E_{\rm R}$, since $\eta \propto s E_{\rm R} $.

For the half-pinned and fully-pinned vortex lattices, there is no
second order contribution to the gap. This is because in these
phases the vortices are located on minima and saddle points of the
pinning potential. The fact that the energy gap in the half-pinned
phases is zero, relies on the fact that we consider infinite
vortex lattices. In a trapped system there will be a gap to
collective excitations, that becomes smaller with increasing
system size.

In Fig. (\ref{fig:gap_vs_V}) we display the gap as function of the
dimensionless parameter $\frac{s E_{\rm R}}{\mu}$ for the cases
$\nu=\frac{1}{4}$ and $\nu=\frac{1}{5}$ which we have considered
in this paper. It is clearly seen that the gap has a discontinuity
if $\frac{s E_{\rm R}}{\mu}$ is tuned through a phase boundary. It
must be emphasized that in the Abrikosov phase only the
qualitative features of the behavior of the gap can be deduced
from Fig. (\ref{fig:gap_vs_V}). This is because the gap then
depends on the shear modulus $\mu_H$ which is a phenomenological
constant in our calculations.

Finally, we would like to comment on the experimental implications
of the collective mode spectra we have calculated. Although our
calculations contain phenomenological parameters, there are
nonetheless some qualitative predictions that could be tested
experimentally. First of all, for the half-pinned and fully-pinned
vortex lattices the collective mode spectrum is anisotropic, i.e.,
the frequency of the collective modes depends on the direction of
propagation. By performing the same experiment as Coddington {\it
et al.} \cite{coddington2003}, in which the collective modes are
excited by a perturbation at the center of the condensate, one
could probe this anisotropy. Another prediction of our theory is
that the collective modes are in general gapped in the presence of
the vortex lattice. An exception is the triangular half-pinned
vortex lattice, which has a gapless mode, corresponding to the
translation of the vortices in one direction. (For the
illustration in Fig.~\ref{fig:fig2} this direction is the
$x$-direction.) For the propagation of the modes in the other
direction there will be a gap. This strong anisotropy should be
experimentally observable by the above-mentioned experiment. The
excitations of the fully-pinned vortex lattice are also gapped. As
mentioned before, the zero-momentum mode corresponds in this case
to a simultaneous in-phase precession of all the vortices around
the maxima of the optical potential. This mode could be excited by
slightly displacing the optical lattice. Because this
zero-momentum mode of the vortex lattice does not have shear or
compression, it does not depend on the elasticity constants, which
are phenomenological in our calculation. Hence
Eq.~(\ref{eq:resultsgap}) should give an accurate prediction for
the gap in this case, which is directly experimentally verifiable.

\section{Conclusions} \label{sec:concl}
In this paper, we have presented a method to determine the ground
state phase diagram of vortices in a Bose-Einstein condensate in
an optical lattice, thereby extending previous work
\cite{reijnders2004} to an arbitrary number of vortices per unit
cell of the optical lattice, the so-called filling factor. The
vortices arrange themselves in various patterns, depending filling
factor and the optical lattice strength. Generally, we find three
vortex phases, \emph{i}) a fully-pinned phase in which each vortex
is pinned to a maximum of the periodic potential, \emph{ii}) a
phase in which half of the vortices are pinned to maxima of the
optical lattice and \emph{iii}) a phase in which none of the
vortices are pinned, and the structure of the vortex lattice is
determined by the interactions. The structure of the unpinned
phase is always hexagonal, whereas the structure of the
half-pinned phase is triangular. We have discussed several
distinct filling factors explicitly, to demonstrate the above
generic features. We calculated the dispersion of low-lying phonon
modes of the vortex lattice for each of these phases. In the case
of the half-pinned and fully-pinned vortex lattice we find that
the collective mode spectrum is anisotropic. Furthermore, in the
unpinned and fully-pinned phase the collective modes are gapped.
Both features should be observable experimentally, and we have
outlined possible experiments to probe the collective modes.

There are several interesting directions for further investigation
of the influence of a periodic potential on the physics of
rotating Bose-Einstein condensates. For instance, it would be
interesting to consider more strongly-correlated regimes that
occur at fast rotation \cite{polini2005}, and to study the effects
of the periodic optical potential on the melting of the vortex
lattice \cite{sinova2002}. One would expect that in this regime
the effect of quantum fluctuations, i.e., quantum tunnelling of
the vortices through the potential barriers of the pinning
potential, becomes important. Aspects of this were studied by S\o
rensen \emph{et al.} \cite{sorensen2004}, who showed that for
ultra low particle and vortex density the ground state of rotating
bosons in a periodic potential is a Laughlin liquid. It would be
challenging to investigate the system with high particle and
vortex density and a large number of vortices per boson.

Yet another interesting possibility for future work is to study a
rotating spinor condensate in the presence of a periodic
potential. Rotating spinor condensates are expected to form
spin-textures (skyrmions) \cite{alkhawaja2001,ruostekoski2001} and regular
lattices thereof \cite{kita2002,reijnders2004_1}, analogous to the
formation of vortices in a single component condensate. The
pinning effects in each spin component of the condensate caused by
the periodic potential will further enrich the phase diagram in
these systems.

We would like to thank M. Hafezi, R. Hagemans, F.J.M. van
Lankvelt, A.H. MacDonald, P. Pedri, V. Schweikhard, K. Schoutens,
and H.T.C. Stoof for useful discussions. This research was
supported by the National Science Foundation under grant
DMR-0115947 (R.A.D.) and by the Netherlands Organization for
Scientific Research, NWO (J.W.R.).


\begin{thebibliography}{99}
\bibitem{footnote2} Apart from breaking the translational symmetry, such a
triangular vortex lattice also breaks a discrete symmetry, because
there are several distinct possibilities to form the triangular
lattice configuration. In this particular case the discrete
symmetry is $Z_4$. This additional symmetry breaking implies the
existence of domain walls between triangular vortex lattices that
are oriented differently. We do not consider such configurations
here.
\bibitem{footnote3} Note that if we rotate the vortex lattice over
an angle $\pi/4$, we we would find the the dynamical matrix is
proportional to $\delta_{\alpha,x}$ instead of
$\delta_{\alpha,y}$. (See also \cite{footnote2}.)
\bibitem{footnote5} We obtained this
estimate using the expression for the shear modulus following from
hydrodynamic theory. We have $e_{\rm shear}=\mu_H A = \mu^{\rm
3D}_H d_z A =\frac{n\hbar\Omega}{8} d_z \pi \ell^2$, with
$\ell^2=\frac{\hbar}{M \Omega}$. The ratio between the chemical
potential $\mu=g n $ and $e_{\rm shear}$ is constant,
$\frac{e_{\rm shear}}{\mu} =\frac{d_z}{32 a_s}$. We then use
$a_s\approx 5$ nm for $^{87}$Rb and $d_z\approx 10 ~\mu {\rm m}$
to find $\eta=\frac{s E_{\rm R}}{e_{\rm shear}}=0.64 \left[\frac{s
E_{\rm R}}{\mu}\right]$.
\end{thebibliography}
\end{document}